\documentclass[12pt,preprint]{aastex}
\usepackage{epsfig}
\usepackage{psfig}

\newcommand{\Chandra}{{\sl Chandra}}
\newcommand{\ROSAT}{{\sl ROSAT}}

\slugcomment{DRAFT 2006.09.19: \apj \, to be submitted}
\shorttitle{{Chandra Search for X-ray Emission from GD~356}}
\shortauthors{M.~C. Weisskopf et al.}

\begin{document}

\title{{A \Chandra\ Search for Coronal X Rays\\ from the Cool White Dwarf GD~356}}

\author{
Martin~C.~Weisskopf\altaffilmark{1},
Kinwah Wu\altaffilmark{2},
Virginia Trimble\altaffilmark{3,4},
Stephen L. O'Dell\altaffilmark{5},\\
Ronald F. Elsner\altaffilmark{5},
Vyacheslav E. Zavlin\altaffilmark{5,6}, and
Chryssa Kouveliotou\altaffilmark{5}
}

\altaffiltext{1}
{NASA Marshall Space Flight Center, VP60, Huntsville, AL 35812 (martin@smoker.msfc.nasa.gov)}
\altaffiltext{2}
{Mullard Space Science Laboratory, University College London, Holmbury St.\ Mary, Surrey RH5~6NT, UK}
\altaffiltext{3}
{Dept.\ of Physics and Astronomy, University of California, Irvine, CA 92697-4575}
\altaffiltext{4}
{Las Cumbres Observatory, Goleta, CA 93117}
\altaffiltext{5}
{NASA Marshall Space Flight Center, VP62, Huntsville, AL 35812}
\altaffiltext{6}
{Research Fellow, NASA Postdoctoral Program (NPP)}

\begin{abstract}
We report observations with the \Chandra\ X-ray Observatory of the single, cool, magnetic white dwarf GD~356.
For consistent comparison with other X-ray observations of single white dwarfs, we also re-analyzed archival \ROSAT\ data for GD~356 (GJ~1205), G~99-47 (GR~290 = V1201~Ori), GD~90, G~195-19 (EG250 = GJ~339.1), and WD~2316+123 and archival \Chandra\ data for LHS~1038 (GJ~1004) and GD~358 (V777~Her).
Our \Chandra\ observation detected no X rays from GD~356, setting the most restrictive upper limit to the X-ray luminosity from any cool white dwarf~---  $L_{X} < 6.0\!\times\! 10^{25}$ erg~s$^{-1}$, at $99.7$\% confidence, for a 1-keV thermal-bremsstrahlung spectrum.
The corresponding limit to the electron density is $n_{0} < 4.4\!\times\!10^{11}$ cm$^{-3}$.
Our re-analysis of the archival data confirmed the non-detections reported by the original investigators.
We discuss the implications of our and prior observations on models for coronal emission from white dwarfs.
For magnetic white dwarfs, we emphasize the more stringent constraints imposed by cyclotron radiation.
In addition, we describe (in an appendix) a statistical methodology for detecting a source and for constraining the strength of a source, which applies even when the number of source or background events is small.
\end{abstract}

\keywords{X rays: individual (GD~356)~--- white dwarfs~--- stars: coronae~--- radiation mechanisms: thermal~--- methods: statistical}

\section{Introduction\label{s:intro}}

Several theorists \citep[e.g., ][]{zhe84,ser90,tho95} have suggested that single, cool, magnetic white dwarfs might have coronae.
Some observers \citep{fon82,arn92,cav93,mus95,mus03} have previously searched for X radiation that might be emitted by hot gas above a white-dwarf photosphere.
There were no persuasive detections, despite one false alarm~--- GR~290, in archival Einstein data \citep{arn92}.
We here report another upper limit, more stringent than the previous ones, for the white-dwarf star GD~356.

Three considerations motivated these X-ray searches: (a) the preponderance of magnetic white dwarfs among the X-ray emitting cataclysmic variables; (b) the possibility that some features in the optical spectra of magnetic white dwarfs might be cyclotron resonances \citep{zhe84}; and (c) the feeling that coronal heating is not so well understood that constraints from other kinds of stars wouldn't be worthwhile.
In the special case of GD~356, the presence of Balmer lines in emission imply at least a chromosphere, making more plausible the existence of a corona.
Magnetic fields in most of the stars examined so far (including GD~356) are 1~MG or more, thought to be fossils from the white dwarfs' previous lives as Ap stars.
More recent measurements of variable, weaker fields in some white dwarfs~--- the DBV GD~358 \citep{win94} and the DA LHS~1038 \citep{sch95}~--- suggest non-fossil magnetism.
Calculations \citep{tho95} indicate that cool white dwarfs with convective envelopes can support $\alpha$--$\omega$ dynamos capable of generating 100-kG fields.

Section~\ref{s:coronal_activity} reviews recent theoretical considerations;
Section~\ref{s:previous_obs}, limits from previous searches.
Next, Section~\ref{s:chandra_obs} presents the results of our \Chandra\ observations of GD~356; 
Section~\ref{s:reanalyze}, our re-analysis of prior \ROSAT\ and \Chandra\ observations of GD~356 and of other white dwarfs.
Finally, Section~\ref{s:discussion} discusses the implications of the new and re-analyzed results and emphasizes the importance of thermal cyclotron radiation in magnetic white dwarfs \citep[cf.][and references therein]{zhe04}.

\section{Magnetic coronal activity from white dwarfs with a convective layer\label{s:coronal_activity}} 

White dwarfs are the class of compact objects that we presumably understand best.
Nevertheless, several fundamental issues remain unresolved.
For instance, are white-dwarf magnetic fields all relic (fossil) or are there transient field components generated by dynamos?
What excites the mysterious line emission from some white dwarfs?
Are X rays from white dwarfs, when detected, simply thermal emission from a deep photosphere? 

All stars show a certain degree of magnetism.
Quite often, coronal activity in non-degenerate stars demonstrates the presence of a magnetic field, presumably produced via magnetohydrodynamic (MHD) processes.
In the canonical model, an $\alpha$-$\omega$ dynamo in a convective stellar envelope generates the field; acoustic waves emerging from deep beneath the stellar atmosphere heat the corona.
In contrast, the magnetic fields of degenerate stars~--- such as white dwarfs~--- may be fossil fields.
Such a fossil field, together with a small, static atmospheric scale height, would lead one to conclude that white dwarfs should not have coronae.
However, some studies have challenged this conclusion.
For example, theoretical arguments \citep{wes80,koe02} suggest that the cooler ($T_{\rm eff}<$ 18,000~K) DA white dwarfs and ($T_{\rm eff}<$ 30,000~K) DB white dwarfs could possess a convective zone~--- a necessary ingredient for coronal X-ray emission.
Further, differential rotation, if it occurs in white dwarfs \citep[cf.][]{kaw99}, might support a magnetic dynamo.
As noted above, the magnetic-field strength generated by an $\alpha$--$\omega$ dynamo in a cool white dwarf might approach 100~kG \citep{tho95}.

Most important is the presence of emission lines from some white dwarfs with no detectable companion.
Among these, and thus especially interesting, are the H-Balmer line emission from the nearby (21.1~pc) white dwarf GD~356 \citep{gre85} and the metal lines from G227-5 and from G35-26 \citep{pro05}.
For single white dwarfs, such emission lines indicate some chromospheric activity.
Alternative models for the H emission lines in GD~356 include accretion of the interstellar medium and the presence of an interacting companion star or even a planet \citep{gre85,lij98}.
The estimated luminosity of the Zeeman-split H$\alpha$ emission lines of GD~356 is $\approx\! 2.1\,\times\, 10^{27}$~erg~s$^{-1}$. 
Further, the flat Balmer decrement \citep[f(H$\alpha$)/f(H$\beta$)~$\approx 1.2$,][]{gre85} excludes photo-ionization and recombination in an optically thin gas, suggesting instead a dense emission region with electron number density $n_{\rm e} \approx 10^{14}$~cm$^{-3}$ \citep{gre85}.
A possible cause of the chromospheric Balmer line emission is irradiation by UV/X rays from a hot magnetic corona above the cooler white-dwarf atmosphere.
The presence of a white-dwarf magnetic corona remains unverified, which motivated our \Chandra\ observation.
However, reported temporal variations of the magnetic field of (DBV) GD~358 \citep{win94} and of (DA) LHS~1038 \citep{sch95} are consistent with a magnetic corona.

In addition to facilitating the MHD dynamo process, a convective zone \citep{boh71} also results in acoustic-wave generation.
Calculations show that the flux of such acoustic waves can be as large as $\approx\! 10^{10}$~erg~cm$^{-2}$ s$^{-1}$ \citep{mus87}.
If the acoustic energy reaches the white-dwarf surface unabsorbed, it can provide a total luminosity $\approx\!5\! \times\! 10^{28}$~erg~s$^{-1}$.
However, due to radiative damping and wave trapping, a substantial fraction of this acoustic energy will not reach the white-dwarf atmosphere.
For cool DA and DB white dwarfs with sensible parameters, fast-mode acoustic waves would attenuate to a negligible level at the white-dwarf atmosphere \citep{mus87}, thus failing to power a magnetic corona.
Instead, wave trapping excites p-mode stellar pulsations \citep{mus89}.
In contrast, if the white dwarf is magnetic (but not so much as to suppress convection), transverse slow modes can propagate, carrying perhaps 60\%--80\% \citep{mus87} of the wave energy into the stellar atmosphere.

\section{Previous X-ray observations\label{s:previous_obs}} 

The detection of X radiation from Sirius~B \citep{mew75} established that white dwarfs are soft-X-ray sources.
Subsequent {\sl Einstein}, {\sl EXOSAT} and \ROSAT\ observations \citep[e.g.,][]{mus87,kah84,pet86,pae89,koe90,kid92,bar93} detected X radiation from a number of hot ($T_{\rm eff}>$ 30,000~K) white dwarfs.
Thermal emission from the photosphere \citep{shi76} accounts for the X-ray emission in all but one case:
For the (DO) KPD~0005+5106, optically thin thermal emission better fits the observed \ROSAT\ data \citep{fle93}, indicating a hot tenuous plasma layer enveloping the white dwarf.
For KPD~0005+5106, the inferred temperature of the X-ray emitting plasma is 0.2--0.3~MK, lower than the typical temperature of magnetic coronae around late-type stars and perhaps indicating a hot wind resembling those of O/B stars.

Searches, with {\sl Einstein} and {\sl EXOSAT} \citep{fon82,arn92} and with \ROSAT\ \citep{cav93,mus95}, for coronal X rays from cool white dwarfs yielded no firm detections.
For searches with the \ROSAT\ Position-Sensitive Proportional Counter (PSPC), \cite{cav93} established X-ray upper limits for the white dwarfs G~99-47 and G~195-19; \citet{mus95}, for GD~90, GD~356, and WD~2316+123.
Likewise, observations \citep{mus03} with the \Chandra\ Advanced CCD Imaging Spectrometer (ACIS) of reportedly magnetically-varying white dwarfs, detected X radiation neither from LHS~1038 nor from GD~358.
Table~\ref{t:prop} summarizes some relevant general properties of these  white dwarfs:
Column~1 gives the name; columns~2 and 3, the epoch-2000 right ascension RA(J2000) and declination Dec(J2000); columns~4 and 5, the proper-motion components; column~6, the distance; column~7, the spectral type; and columns~8 and 9, the surface effective temperature and magnetic field.

The previous searches determined upper limits to the X-ray luminosity for a thermal-bremsstrahlung spectrum with negligible interstellar absorption.
Assuming a coronal temperature of 11.6~MK (1.0~keV), \citet{cav93} obtained \ROSAT-PSPC 99\%-confidence limits to the 0.1--2.5-keV luminosity of $<2.9\! \times\! 10^{26}$~erg~s$^{-1}$ for G~99-47 and $<1.4\! \times\! 10^{26}$~erg~s$^{-1}$ for G~195-19.
Assuming a coronal temperature of 2.5~MK (0.215~keV), \citet{mus95} obtained \ROSAT-PSPC 99.7-\%-confidence (3-$\sigma$) limits to the 0.1--2.4-keV  luminosity of $<7.8\! \times\! 10^{27}$~erg~s$^{-1}$ for GD~90, $<4.4\! \times\! 10^{26}$~erg~s$^{-1}$ for GD~356, and $<3.4\! \times\! 10^{27}$~erg~s$^{-1}$ for WD~2316+123.
Under the same assumptions but using \Chandra-ACIS observations, \citet{mus03} found $<4.3\! \times\! 10^{26}$~erg~s$^{-1}$ for LHS~1038 and $<4.3\! \times\! 10^{27}$~erg~s$^{-1}$ for GD~358.
These upper limits were comparable to the X-ray luminosities that \citet{mus03} had predicted for LHS~1038 and GD~358~--- $5\! \times\! 10^{26}$~erg~s$^{-1}$ and $5\!\times\!10^{27}$~erg~s$^{-1}$, respectively.
For uniform comparison of results over a wider range of assumed parameters, we re-analyzed (\S\ref{s:reanalyze} and Table~\ref{t:x-data}) the relevant archived \Chandra\ (\S \ref{s:chandra}) and \ROSAT\ (\S \ref{s:rosat}) data.

\section{The \Chandra\ observation of GD~356\label{s:chandra_obs}}

We obtained a 31.8-ks \Chandra\ observation (ObsID 4484, 2005 May 24) using the Advanced CCD Imaging Spectrometer (ACIS) S3 (back-illuminated) CCD in the faint, timed-exposure mode, with 3.141-s frame time.
Background levels were nominal throughout the observation.
Standard \Chandra\ X-ray Center (CXC) processing (ASCDS version number CIAO3.2) provided level-2 event files.
In analyzing data, we utilized events in pulse-invariant channels corresponding to 0.5 to 8.0 keV.

Our \Chandra\ observation found no X rays in a $1\arcsec$-radius detect cell at the epoch-2005.39 position of GD~356~--- $16^{\rm h}\,40^{\rm m}\,57\fs23$ $+53\degr\, 41\arcmin\, 8\farcs6$, after adjustment of the epoch-2000 coordinates for proper motion (Table~\ref{t:prop}).
The nearest detected X-ray source~--- \#6 in Table~\ref{t:field_gd356}~--- lies 46\arcsec\ from GD~356.
To determine the background,  we ``punched out'' the detected X-ray sources eliminating all counts within a 20-$\sigma$ radius of each position in Table~\ref{t:field_gd356}.
Table~\ref{t:x-data} summarizes the results for our \Chandra\ observation of GD~356, as well as for our re-analysis (\S \ref{s:reanalyze}) of previous \Chandra\ (\S \ref{s:chandra}) and \ROSAT\ (\S \ref{s:rosat}) searches for X-ray emission from this and other single white dwarfs.
Column~1 lists the name of the white dwarf; column~2, the observatory used; column~3, the epoch relative to 2000.00 (used to adjust for proper motion); and column~4, the integration time.
Column~5 presents the number of detected counts $m_{T}$ in the ``Target'' (detect) cell $T$ and column~6, its solid angle $\Omega_{T}$.
Analogously, column~7 presents the number of detected counts $m_{R}$ in the ``Reference'' region $R$ and column~8, its solid angle $\Omega_{R}$.
Finally, columns~9, 10, 11, and 12 list the confidence level for detection of a source (\S \ref{s:detect}), the 99.7\%-confidence upper limit on the expectation value of the number of source counts (\S \ref{s:limit}), and the corresponding limits on the X-ray luminosity (for a 1-keV thermal-bremsstrahlung spectrum) and on the electron density, which follow from the analysis below.
Owing to the very small (1-$\arcsec$ radius) detect cell afforded by \Chandra's sub-arcsecond resolution, only two (2) counts in the Target aperture would have constituted a (3-$\sigma$) 99.7\%-confidence detection for either of the three \Chandra\ data sets!
Neither of the three \Chandra\ observations found an event in the detect cell.

A statistical analysis (Appendix~\ref{s:stat}) of the data yields (Equation~\ref{e:src-conf}) the confidence level for detection of a source within the Target (detect) cell and (Equation~\ref{e:conf-s}) a 99.7\%-confidence upper limit to the expected number of source counts in the Target (detection) cell $T$.
Before converting the source counts to a flux and luminosity, we correct for the fraction of the point spread function (PSF) outside the detect cell.
For \Chandra\ and almost any soft spectrum, about 90\%\footnote{see http://asc.harvard.edu/proposer/POG/index.html} of the source photons for an on-axis source would lie within the chosen detect cell.
To convert the resulting 5-count (4.5/0.9) limit to a flux, we calculated the redistribution matrix (rmf) and effective area (arf) functions appropriate to the location of GD~356 in the focal plane, using the \Chandra\ CIAO 3.3 software tools {\tt mkacisarmf} and {\tt mkarf}, following the analysis thread\footnote{see http://cxc.harvard.edu/ciao/ahelp/mkacisrmf.html} for creating these functions for a specific location.
Assuming a column $N_{H} =5\!\times\!10^{18}$~cm$^{-2}$ and a thermal bremsstrahlung spectrum, we used XSPEC (v11.3.2)\footnote{see http://xspec.gsfc.nasa.gov/} with {\tt abund} set to {\tt wilm}, {\tt xsect} set to {\tt vern},  and {\tt tbabs(bremss)} as the model.
For each assumed value for the coronal temperature, we adjusted the model normalization until the absorbed flux produced 5.0 counts in the S3 detector in the observing time of the \Chandra\ observation.
Feeding this normalization into XSPEC, we used the {\tt dummyrsp} feature to calculate the flux for $10^4$ bins over the energy range $10^{-5}$~keV to 100~keV.
We used this approach to improve the accuracy of our flux calculations since the instrument response is calculated over a more restricted energy range, and with cruder energy bins.
We then calculated fluxes over the band from 0.01 to 100~keV, now setting the column density to zero to obtain the unabsorbed flux.

The upper panel of Figure~\ref{f:luminosity} plots the 3-$\sigma$ upper limit to the total luminosity, as a function of coronal temperature, for our observation of GD 356 and for our re-analysis (\S \ref{s:reanalyze}) of previous white-dwarf observations.
The upper panel of Figure~\ref{f:density} shows the corresponding upper limit to the electron density for an effective emitting volume $4 \pi R^2 H$~--- appropriate for a geometrically thin ($H \ll R$), transparent, atmosphere of (exponential) scale height $H$ around a sphere of radius $R$.
For the ion density, we assume a fully ionized plasma of hydrogen and helium with $n_{\rm He}/n_{\rm H}=0.1$, such that $\sum{n_{i} Z_{i}^2} = 1.4\, n_{e}$.
The bremsstrahlung emissivity then goes as $1.4\, \overline{n_{e}^2} = 1.4\, n_{0}^2 / 2$, where $n_{0}$ is the electron density at the base of an isothermal corona.
Note that only about half the coronal emission emerges, due to photospheric absorption of most of the downward coronal flux.
Using GD~356's 21.1-pc parallax distance \citep{van95} and published UBV \citep{mer94} and JHK \citep{skr06} photometry, we obtain the photospheric temperature $T_{s} = 7840$~K and radius $R = 0.0105\, R_\odot = 7.34\!\times\! 10^{8}$~cm.
For the coronal scale height, we use $H = (2\!\times\!10^{7}$ cm) $kT$/(1~keV), corresponding to a surface gravity log[$g$(cgs)] = 8.
The lower panels of Figures~\ref{f:luminosity} and \ref{f:density} display analogous limits for the {\tt tbabs(mekal)} Mewe--Kaastra--Liedahl (MEKAL) model with parameters as above, but {\tt abund} set to {\tt lodd}.

Besides examining the white-dwarf location for X rays, we searched for X-ray sources anywhere on S3 employing techniques described in \citet{ten06}.
For the 31.8-ks \Chandra\ observation, Table~\ref{t:field_gd356} lists the X-ray properties of the 23 detected sources, each designated with a source number in column~1.
Columns~2--5 give, respectively, right ascension RA(J2000), declination Dec(J2000), extraction radius $\theta_{\rm ext}$, and approximate number of X-ray counts $m_{D}$ detected from the source.
The single-axis RMS error in the X-ray-source position is $\sigma_{X}=[(\sigma_{\rm PSF}^2/m_{D}) + \sigma^2_{\rm sys}]^{1/2}$, where $m_{D}$ is the approximate number of detected events above background, $\sigma_{\rm PSF}$ is the dispersion of the circular Gaussian that approximately matches the PSF at the source location, and $\sigma_{\rm sys}$ is a systematic error.
Uncertainties in the plate scale\footnote{See
http://asc.harvard.edu/cal/hrma/optaxis/platescale/} imply $\sigma_{\rm sys}\approx 0\farcs13$: 
To be conservative, we set $\sigma_{\rm sys}=0\farcs2$ (per axis).
Column~6 gives the radial uncertainty $\theta_{99}=3.03\: \sigma_{X}$ in the X-ray position~--- i.e., $\chi^{2}_{2}=9.21=3.03^2$ corresponds to 99\% confidence on 2 degrees of freedom, for inclusion of the true source position.

\section{Re-analysis of previous observations.\label{s:reanalyze}}

For consistent comparison of X-ray observations, we also re-analyzed certain prior \Chandra\ (\S \ref{s:chandra}) and \ROSAT\ (\S \ref{s:rosat}) observations of single, cool white dwarfs.
The \Chandra\ observations used the ACIS-S instrument; the \ROSAT\ observations, the PSPC.

\subsection{Prior \Chandra\ Observations\label{s:chandra}}

We re-processed and analyzed previous \Chandra\ observations of the cool white dwarfs LHS~1038 (ObsID~1864, 5.88~ks) and GD~358 (ObsID~1865, 4.88~ks), in the same manner as for GD~356 (\S\ref{s:chandra_obs}).
Table~\ref{t:x-data} summarizes relevant parameters and results.
Figure~\ref{f:luminosity} shows the total luminosity (3-$\sigma$) 99.7\%-confidence limits, for distances listed in Table~\ref{t:prop}.
Figure~\ref{f:density} plots the corresponding upper limits to the electron density, approximating the radius of each white dwarf by that of GD~356 (\S \ref{s:chandra_obs}).

As described above for our observation of the GD-356 field (\S \ref{s:chandra_obs}), we also searched for X-ray sources in the S3 observations of the LHS-1038 and GD-358 fields.
In the same form as Table~\ref{t:field_gd356}, Tables~\ref{t:field_lhs1038} and \ref{t:field_gd358} list X-ray sources detected on S3, for the LHS-1038 and GD-358 fields, respectively.

\subsection{\ROSAT\ Observations\label{s:rosat}}

Table~\ref{t:x-data} also summarizes our re-analysis of the \ROSAT-PSPC observations, first analyzed by \citet{cav93} and by \citet{mus95}.
For the re-analysis, we selected the 0.1--2.4-keV energy band, a Target region of 1.5\arcmin\ (source extraction) radius, and an annular Reference region 1.5\arcmin--2.5\arcmin\ radius (for background estimation), as did  \citet{mus95}.
For sufficiently soft sources ($E < 0.5$ keV), most ($\ge 97\%$) of the flux from a source lies within the 1.5\arcmin-radius Target (detect) region \citep{boe00}.
Toward higher energies, the upper limits will be somewhat conservative because a decreasing fraction of source events appears in the Target aperture, while an increasing fraction appears in the Reference aperture.
Furthermore, the \ROSAT\ exposures are not sufficiently long to measure the background accurately.
Appendix~A describes a statistical methodology for dealing with these  issues.

Following procedures described above (\S \ref{s:chandra_obs}) but now using the \ROSAT-PSPC response matrix, we established (3-$\sigma$) 99.7\%-confidence upper limits to the expected number of source counts and then converted those to flux and luminosity.
For calculating luminosity, we used the distances listed in Table~\ref{t:prop}.
Figure~\ref{f:luminosity} plots the derived luminosity limits as a function of assumed temperature for the bremsstrahlung (upper panel) and MEKAL (lower panel) spectral models.
Figure~\ref{f:density} shows the corresponding 3-$\sigma$ upper limits to the electron density, again approximating the radius of each white dwarf by  that of GD~356 (\S \ref{s:chandra_obs}).

\section{Discussion\label{s:discussion}}

The objective of this investigation was to determine whether single, cool white dwarfs emit X radiation indicative of magnetic coronal activity.
To assess the scientific implications of these null detections, we first (\S \ref{s:formation}) discuss issues related to the formation of a magnetic corona around a white dwarf.
We then (\S \ref{s:cyclotron}) examine the rather severe constraints that cyclotron emission lines and radiative loss impose on a corona around a cool magnetic white dwarf, such as GD~356.
Finally (\S \ref{s:conclude}), we summarize our results and conclusions about hypothesized hot coronae around magnetic white dwarfs.

\subsection{Formation of a Corona\label{s:formation}}

The existence of coronae around single, cool white dwarfs remains an unsettled issue: This, of course, motivated our search for X-ray evidence.
Here we briefly address some questions relevant to the formation of a magnetic corona.
\begin{enumerate}
\item Is there a convection zone?
Theoretical studies have indicated that convection can occur in white dwarfs for certain temperature ranges.
Analyses of white-dwarf atmospheric elemental abundance also evidence convective activity.
Thus, it is quite plausible that some white dwarfs possess a convection zone that generates acoustic waves.
\item How much acoustic wave energy is generated in the convection zone?
Calculations \citep{mus87,win94} have shown that the convection-generated acoustic flux could exceed $10^8$~erg~cm$^{-2}\,$s$^{-1}$ and may reach $10^{10}$~erg~cm$^{-2}\,$s$^{-1}$ for DA white dwarfs and $10^{11}$~erg~cm$^{-2}\,$s$^{-1}$ for DB white dwarfs.
\item What fraction of the acoustic wave energy is transmitted to the white-dwarf surface?
For white dwarf with a substantial magnetic field ($B \approx 10^4$~G or higher), more than half of the wave energy generated in the convection zone can reach the white-dwarf atmosphere.
If all the wave energy is converted into coronal X rays, the expected luminosity will be $\approx 10^{27}\!-\! 10^{30}$~erg~s$^{-1}$.
Our upper limit to the X-ray luminosity for GD~356 is an order of magnitude below $10^{27}$~erg~s$^{-1}$ over most of the range of assumed temperatures.
This implies either that the putative convection zone generates less acoustic energy, that the efficiency of acoustic-wave transmission to the surface is smaller, or that a corona fails to form even if the acoustic energy reaches the white-dwarf surface.
\item Can a magnetic corona form, given sufficient energy provided by the acoustic flux?
The null detection of coronal X rays from white dwarfs has led some \citep [see e.g.,][]{mus05} to suggest that the emerging acoustic flux causes chromospheric activity rather than formation of a hot corona.
An example of such chromospheric activity would be oscillations resulting from the stellar atmosphere to propagating acoustic waves in the presence of a temperature inversion \citep{mus05}.
Whether acoustic waves generate chromospheric activity remains unverified.
Nevertheless, a number of systems exhibit chromospheric activity.
The luminosity of H$\alpha$ emission lines in GD~356 is $\approx 1.8\! \times\! 10^{27}$~erg~s$^{-1}$ \citep[see][]{gre85}.
Our \Chandra\ observation eliminates the possibility that irradiation of the atmosphere by coronal X rays powers the Balmer lines.
Further, it is not clear that atmospheric oscillations can account for the luminosity of Balmer emission lines in GD~356.
\end{enumerate}

While coronal X radiation from cool ($T_{\rm eff} < 10,000$~K) magnetic white dwarfs remains undetected, X radiation from hot optically thin thermal plasmas appears to occur from very hot ($1.2 \times 10^5$~K) white dwarf KPD 0005+5106.
Now the question is this:
Without a magnetic corona, what supports the hot plasma envelope in the strong gravitational field of the white dwarf?
Radiation driven envelopes~--- predicted for hot white dwarfs \citep{bes90,zhe96}~--- could provide this support.
Provided the white dwarf has a strong magnetic field, cyclotron-resonance radiation pressure from photospheric radiation could drive the wind, with an estimated mass-loss rate $\approx 2\! \times\! 10^{10}$~g~s$^{-1}$.
Thus, a hot magnetic white dwarf could emit X rays from an optically thin, thermal plasma in a radiation-driven outflow.

\subsection{Cyclotron Radiation\label{s:cyclotron}}

The \Chandra\ observation sets stringent constraints on a supposed hot corona above the white-dwarf atmosphere.
For GD~356, the X-ray luminosity $L_{X} < 6.0\!\times\! 10^{25}$~erg s$^{-1}$ and electron number density $n_{0} < 4.4\!\times\! 10^{11}~{\rm cm}^{-3}$ at the base of a corona of (exponential) scale height of $2.0\!\times\! 10^7~{\rm cm}$ for a 1-keV plasma.
However, for a magnetic white dwarf, \citet{zhe04} note that electron-cyclotron emission lines and radiative losses even more severely constrain the parameters of a hypothesized hot corona.

For the 15-MG field of GD356, we computed the radiative transfer in a hot corona, including cyclotron and bremsstrahlung emissivities and opacities due to cyclotron radiation \citep{cha89} and Faraday rotation and mixing.
Figure~\ref{f:cyclotron} displays the emergent spectrum for white-dwarf parameters appropriate to GD~356 (\S \ref{s:chandra_obs}) and coronal density and temperatures given in the caption.
For a density comparable to the upper limit set by the \Chandra\ observation, the hypothesized hot corona would emit extremely strong thermal cyclotron lines in the near-infrared band~--- namely, around the second and third harmonics of the cyclotron frequency ($\nu_{B} = 42$~THz, $\lambda_{B} = 7.1$~$\mu$m).
The JHK photometry \citep{skr06} shows no indication of such an excess, indicating that a hot corona does not exist or has a density much lower than the upper limit set by the X-ray observations.
Alternatively, the magnetic field could be somewhat weaker than 15~MG, which would shift the strong third harmonic ($\nu_{3} = 126$~THz, $\lambda_{3} = 2.4$~$\mu$m) to a lower frequency, longward of the K$_{s}$ band.
Nevertheless, our calculation confirms the conclusion of \citet{zhe04}:
Infrared--visible spectrophotometry is potentially a powerful probe of any hot corona around a magnetic white dwarf.

In addition to producing potentially detectable emission lines, thermal cyclotron radiation would be the dominant cooling mechanism in a hot corona around a magnetic white dwarf.
Figure~\ref{f:power} displays the radiated thermal cyclotron and bremsstrahlung luminosity for a supposed isothermal corona around GD~356 ($B = 15$~MG), as a function of density for various electron temperatures between 0.125~keV and 2.0~keV.
As the plot clearly shows, cyclotron cooling dramatically exceeds bremsstrahlung cooling for a hot, tenuous plasma above a magnetic white dwarf.
Indeed, for GD~356, the coronal thermal cyclotron luminosity rivals the photospheric luminosity unless the electron density is very much less than the upper limit set by the X-ray observation.

Strong cyclotron cooling above a magnetic white dwarf imposes severe demands upon any coronal heating mechanism.
As Figure~\ref{f:power} demonstrates, even the weak requirement \citep{zhe04} that the coronal thermal cyclotron luminosity not exceed the photospheric luminosity limits the electron density to $n_{0} < 3\!\times\! 10^{9}~{\rm cm}^{-3}$ for $kT = 1$~keV.
If an acoustic flux as large as $10^{10}$~erg~cm$^{-2}$ s$^{-1}$ \citep{mus87} efficiently heats a corona around GD~356, this heating rate can balance cyclotron cooling for $n_{0} \approx 3\!\times\! 10^{5}~{\rm cm}^{-3}$ for $kT = 1$~keV.
This is about 6-orders-of-magnitude less than the density limit set by X-ray non-detection!

For an electron density $n_{0} < 3\!\times\! 10^{6}~{\rm cm}^{-3}$~--- approximately independent of temperature~---
a corona above the white-dwarf photosphere would be transparent to cyclotron radiation in the 15-MG field of GD~356.
However, when transparent, the cyclotron cooling time~--- again approximately independent of temperature~--- is only 2~$\mu$s in this magnetic field.
For electron densities and temperatures of interest, the mean time between collisions~--- about $(1.4\!\times\! 10^{9}~{\rm s~cm}^{-3})\, (kT/(1~{\rm keV}))^{1.5}\, {n_{e}}^{-1}$~--- is much longer than this cyclotron cooling time.
Thus, the plasma would not be in local thermodynamic equilibrium \citep[LTE,][and references therein]{zhe04}, unless some collisionless process~--- e.g., scattering by Alfven waves~--- intervenes to transfer energy to the electrons' transverse degrees of freedom, on this very short timescale.

\subsection{Summary\label{s:conclude}}

In summary, our \Chandra\ observation of the single, cool white dwarf GD~356 limits the luminosity and density of a hypothesized hot corona to
$L_{X} < 6.0\!\times\! 10^{25}$ erg~s$^{-1}$ and $n_{0} < 4.4\!\times\!10^{11}$ cm$^{-3}$ ($99.7$\% confidence), for a 1-keV thermal-bremsstrahlung spectrum.
We also re-analyzed archival \ROSAT\ data for this white dwarf, G~99-47, GD~90, G~195-19, and WD~2316+123, as well as archival \Chandra\ data for LHS~1038 and GD~358, using a statistical methodology (described in the Appendix) better suited to the low-count observations.
Our upper limits are reasonably consistent with those of the original authors and lie somewhat above the limits we set for GD~356.
As an aside to these searches for X-ray emission from cool white dwarfs, we have listed all \Chandra-detected sources in the LHS-1038, GD-356, and GD-358 fields on the ACIS-S3 CCD.

We have shown \citep[see also][]{zhe04} that for a magnetic white dwarf (such as GD~356), the non-detection of infrared--visible cyclotron emission lines can more severely constrain the parameters for a hot corona than does an X-ray non-detection.
Furthermore, strong cyclotron cooling places extreme demands upon any coronal-heating mechanism.
Indeed, our preliminary theoretical analysis suggests that cyclotron cooling around a magnetic white dwarf renders problematic the formation and maintenance of a hot corona.
However, we require a more thorough study to prove this conclusively.

\acknowledgments
Those of us at NASA's Marshall Space Flight Center (MSFC) acknowledge support from the \Chandra\ Program.
VEZ acknowledges support from the NASA Postdoctoral Program (NPP).
Our analyses utilized software tools from the \Chandra\ X-ray Center (operated for NASA by the Smithsonian Astrophysical Observatory, Cambridge MA) and from the High-Energy Astrophysics Science Archive Research Center (HEASARC, operated by the NASA Goddard Space Flight Center, Greenbelt MD, and by the Smithsonian Astrophysical Observatory, Cambridge MA) 
In addition, our research utilized the SIMBAD database and VizieR catalog access tool (operated at Centre de Donn\'{e}es astronomiques de Strasbourg, France) and NASA's Astrophysics Data System (operated by the Harvard--Smithsonian Center for Astrophysics, Cambridge MA).

\clearpage
{\appendix
\section{Statistical Methodology \label{s:stat}}

Statistical estimates of source and background counts and their errors often merely approximately describe an observation or apply only in the large-number limit.
Here we present a statistical methodology that more generally describes apertured data (including allowing for uncertainty in the background) and makes no large-number assumption.
First (\S \ref{s:basis}) we obtain the probability distribution that accurately describes the observation and provides the basis for the statistical analyses to follow.
We then describe an appropriate statistical test for detection of a source (\S \ref{s:detect}) and one for constraining the expectation value for the number of source events (\S \ref{s:limit}).

\subsection{Probability for Observed Events \label{s:basis}}

We characterize a measurement (realization) in terms of the observed number of events (counts) $m_T$ and $m_R$ in disjoint regions (apertures) $T$ and $R$ of known measure (solid-angle, area, wavelength band, time interval, etc., as appropriate) $\Omega_{T}$ and $\Omega_{R}$, respectively.
We regard $T$ as a ``Target'' aperture that {\em may} contain a source with an expectation value $\bar{m}_{S}$  events (counts); $R$ as a ``Reference'' aperture that contains {\em no} source.
Although no source lies within $R$, source events may occur in $R$ if their distribution is not delta-distributed~--- i.e., confined to a point.
Thus, we define $\Psi_{T}$ and $\Psi_{R}$ to be the known fractions of source events in the Target $T$ and Reference $R$ apertures, such that the expectation values for the number of source events are $\bar{m}_{S} \Psi_{T}$ and $\bar{m}_{S} \Psi_{R}$, respectively.
In addition to source events, apertures $T$ and $R$ contain background (non-source) events, with expectation values $\bar{\mu}_B \Omega_{T}$ and $\bar{\mu}_B \Omega_{R}$, with $\bar{\mu}_B$ the expectation value for the density (per unit measure) of background events.
For convenience, we denote with a subscripted ``$U$'' parameters or values over the combined aperture $U \equiv T \cup R$, with $T \cap R = 0$~---
namely, $\Omega_{U} = \Omega_{T}+\Omega_{R}$, $\Psi_{U} = \Psi_{T}+\Psi_{R}$, and $m_{U} = m_{T}+m_{R}$.

Consequently, the expectation values for the number of events (counts) in apertures $T$ and $R$ are $\bar{m}_{T}$ and $\bar{m_{R}}$, respectively:
\begin{eqnarray}
\bar{m}_{T} & = & \bar{m}_{S} \Psi_{T} + \bar{\mu}_B \Omega_{T} \ , \vspace{1in} \label{e:mt} \\
\bar{m}_{R} & = & \bar{m}_{S} \Psi_{R} + \bar{\mu}_B \Omega_{R} \ . \label{e:mr}
\end{eqnarray}
Hence, the probability for $m_T$ and $m_R$ events in an observation (realization) is
\begin{eqnarray}
 P_{m_{T},m_{R}}(\bar{m}_{T},\bar{m}_{R}) & = & [\, \bar{m}_{T}^{m_T} \ e^{-\bar{m}_{T}}/m_T!\, ] \times [\, \bar{m}_{R}^{m_R} \ e^{-\bar{m}_{R}}/m_R!\, ] \ . \label{e:pmtmr}
\end{eqnarray}
Upon substituting Equations~\ref{e:mt} and \ref{e:mr} into Equation~\ref{e:pmtmr},
\begin{eqnarray}
 P_{m_{T},m_{R}}(\bar{m}_{S},\bar{\mu}_{B}; \Psi_{T},\Psi_{R},\Omega_{T},\Omega_{R}) & = & [\, (\bar{m}_{S} \Psi_{T}+\bar{\mu}_{B} \Omega_{T})^{m_{T}} \ e^{-(\bar{m}_{S} \Psi_{T}+\bar{\mu}_{B} \Omega_{T})}/m_{T}!\, ] \times \nonumber \\
   &   & [\, (\bar{m}_{S} \Psi_{R}+\bar{\mu}_{B} \Omega_{R})^{m_{R}} \ e^{-(\bar{m}_{S} \Psi_{R}+\bar{\mu}_{B} \Omega_{R})}/m_{R}!\, ] \ . \label{e:pmsub}
\end{eqnarray}
Given values for the known parameters ($\Psi_{T}$, $\Psi_{R}$, $\Omega_{T}$, and $\Omega_{R}$) and for the observed number of events ($m_{T}$ and $m_{R}$) in each aperture, Equation~\ref{e:pmsub} provides the basis for statistical tests to constrain expectation values for source events ($\bar{m}_{S}$) and for background event density ($\bar{\mu}_{B}$).

\subsection{Detection of a Source \label{s:detect}}

The first type of statistical test addresses detection.
Note that this is a test for {\em detection} only:
It provides neither a measured value nor an upper limit.
In order to test for detection of a source in the target aperture $T$, we investigate the hypothesis that there is {\em no source}~--- i.e., that $\bar{m}_{S} = 0$.
Under this null hypothesis, the conditional probability of obtaining $m_{T}$ and $m_{R}$ (background) events in apertures $T$ and $R$, given $m_{U} \equiv m_{T} + m_{R}$ events in both apertures, is
\begin{eqnarray}
 P_{m_{T},m_{R}}(0,\bar{\mu}_{B}; \Omega_{T},\Omega_{R}\ |\ m_{U}) & = & P_{m_{T},m_{R}}(0,\bar{\mu}_{B}; \Omega_{T},\Omega_{R})\ /\ P_{m_{U}}(0,\bar{\mu}_{B}; \Omega_{U}) 
\label{e:src-cond} \ ,
\end{eqnarray}
with $\Omega_{U} \equiv \Omega_{T} + \Omega_{R}$.
From the Poisson distribution (Equation~\ref{e:pmsub}), the conditional probability (Equation~\ref{e:src-cond}) reduces to the obvious binomial distribution, independent of $\bar{\mu}_{B}$ under the null hypothesis:
\begin{eqnarray}
 P_{m_{T},m_{R}}(\Omega_{T},\Omega_{R}\ |\ m_{U}; \bar{m}_{S}\!=\!0) & = & P_{m_{T},m_{R}}(\Omega_{T},\Omega_{R}) \ / \ P_{m_{U}}(\Omega_{U}) \nonumber \\
\nonumber & = &  \left(\frac{\Omega_{T}^{m_{T}}}{m_{T}!} \right)  \left(\frac{\Omega_{R}^{m_{R}}}{m_{R}!} \right) \ / \  \left(\frac{\Omega_{U}^{m_{U}}}{m_{U}!} \right) \\
  & = & \frac{m_{U}!}{m_{T}!\, m_{R}!} \, \left( \frac{\Omega_{T}}{\Omega_{U}} \right)^{m_{T}} \left( \frac{\Omega_{R}}{\Omega_{U}} \right)^{m_{R}} .
\label{e:src-disc}
\end{eqnarray}

The cumulative probability of obtaining $m_{T}$ or more events in the Target aperture $T$, given $\bar{m}_{S}\!=\!0$ and $m_{U} = m_{T} + m_{R}$ events in the combined aperture $\Omega_{U} = \Omega_{T} + \Omega_{R}$ is then
\begin{eqnarray}
{\cal P}(\ge m_{T}\ | \ m_{U}; \bar{m}_{S}\!=\!0) & = & \sum_{m=m_{T}}^{m_{U}} \frac{m_{U}!}{m!\, (m_{U}-m)!} \, \left( \frac{\Omega_{T}}{\Omega_{U}} \right)^{m} \left(1 - \frac{\Omega_{T}}{\Omega_{U}} \right)^{m_{U}-m} .
\label{e:src-cumu}
\end{eqnarray}
Consequently, Equation~\ref{e:src-cumu} gives a confidence level ${\cal C}$ for detection of a source~--- i.e., for showing that $\bar{m}_{S} > 0$.
\begin{eqnarray}
{\cal C}(\bar{m}_{S}\!>\!0\ | \ m_{T}, m_{R}; \Omega_{T}, \Omega_{R}) & = & {\cal P}(< m_{T}\ | \ m_{U}; \bar{m}_{S}\!=\!0) \ = \ 1-{\cal P}(\ge\! m_{T}\, |\,  m_{U}; \bar{m}_{S}\!=\!0) \nonumber \\
& = & \sum_{m=0}^{m_{T}-1} \frac{m_{U}!}{m!\, (m_{U}-m)!} \, \left( \frac{\Omega_{T}}{\Omega_{U}} \right)^{m} \left(1 - \frac{\Omega_{T}}{\Omega_{U}} \right)^{m_{U}-m} .
\label{e:src-conf}
\end{eqnarray}
This expression is valid for any number of events, in either the Target or the Reference aperture.
Thus, it does not require that the background event density ($\mu_{B}$) is statistically well determined.

If the expectation value for the background event density is well known, then we can simplify Equation~\ref{e:pmsub} to the more familiar
\begin{eqnarray}
P_{m_{T}}(0,\mu_{B};\Omega_{T}) & = & \frac{(\mu_{B} \Omega_{T})^{m_{T}}}{m_{T}!} \ e^{-\mu_{B} \Omega_{T}} \ . \label{e:pub}
\end{eqnarray}
Given $\mu_{B}$, the corresponding cumulative probability of obtaining $m_{T}$ or more (background) events in the Target aperture then becomes
\begin{eqnarray}
{\cal P}(\ge m_{T}\ | \ \mu_{B} \Omega_{T}; \bar{m}_{S}\!=\!0) & = & \sum_{m=m_{T}}^{\infty} \frac{(\mu_{B} \Omega_{T})^{m}}{m!} \ e^{-\mu_{B} \Omega_{T}} \ . \label{e:pmt-cumu}
\end{eqnarray}
Therefore, Equation~\ref{e:pmt-cumu} yields a confidence level ${\cal C}$ for detection of a source~--- i.e., for showing that $\bar{m}_{S} > 0$.
\begin{eqnarray}
{\cal C}(\bar{m}_{S}\!>\!0\ | \ m_{T}; \mu_{B} \Omega_{T}) & = & {\cal P}(< m_{T}\ | \ \mu_{B} \Omega_{T}; \bar{m}_{S}\!=\!0) \ = \ 1-{\cal P}(\ge m_{T}\ | \ \mu_{B} \Omega_{T}; \bar{m}_{S}\!=\!0) \nonumber \\
& = & \sum_{m=0}^{m_{T}-1} \frac{(\mu_{B} \Omega_{T})^{m}}{m!} \ e^{-\mu_{B} \Omega_{T}} \ .
\label{e:pmt-conf}
\end{eqnarray}

\subsection{Measurement of Source \label{s:limit}}

The second type of statistical test addresses measurement of the expectation value $\bar{m}_{S}$ for the number of source events.
Using Equation~\ref{e:pmsub} as a likelihood function for the parameters $m_{S}$ and $\mu_{B}$, we obtain maximum-likelihood estimators for each.
\begin{eqnarray}
\hat{m}_{S} & = & \frac{m_{T} \Omega_{R} - m_{R} \Omega_{T}}{\Psi_{T} \Omega_{R} - \Psi_{R} \Omega_{T}} \label{e:ms-est} \ , \\ 
\hat{\mu}_{B} & = & \frac{m_{R} \Psi_{T} - m_{T} \Psi_{R}}{\Psi_{T} \Omega_{R} - \Psi_{R} \Omega_{T}} \ . \label{e:ub-est} 
\end{eqnarray}
Evaluation of the second-order partial derivatives of these parameters about their maximum-likelihood estimators leads to estimators for the components of the covariance matrix.
\begin{eqnarray}
\hat{\sigma}_{m_{S}}^2 = {\rm covar}(m_{S},m_{S}) & = & \left( \frac{{\Psi_{T}}^2}{m_{T}} + \frac{{\Psi_{R}}^2}{m_{R}} \right)^{-1} \ , \label{e:ms-var} \\
\hat{\sigma}_{\mu_{B}}^2 = {\rm covar}(\mu_{B},\mu_{B}) & = & \left( \frac{{\Omega_{T}}^2}{m_{T}} + \frac{{\Omega_{R}}^2}{m_{R}} \right)^{-1} \ , \label{e:ub-var} \\
\hat{\sigma}_{m_{S},\mu_{B}} = {\rm covar}(m_{S},\mu_{B}) & = & \left( \frac{\Psi_{T}\Omega_{T}}{m_{T}} + \frac{\Psi_{R}\Omega_{R}}{m_{R}} \right)^{-1} \ . \label{e:co-var}
\end{eqnarray}
Here, $\hat{\sigma}_{m_{S}}$ and $\hat{\sigma}_{\mu_{B}}$ are the maximum-likelihood estimators for the standard deviation in $m_{S}$ and $\mu_{B}$; $\sigma_{m_{S},\mu_{B}} \ne 0$ shows that the estimators for $m_{S}$ and $\mu_{B}$ are correlated.

Equations~\ref{e:ms-var}, \ref{e:ub-var}, and \ref{e:co-var} do not accurately describe the probability distribution for $m_{S}$ and $\mu_{B}$ except in the large-number limit---i.e., when the probability is approximately normally distributed.
Thus, to obtain an accurate description of the probability density function for $m_{S}$ and $\mu_{B}$, we return to Equation~\ref{e:pmsub}.

Equation~\ref{e:pmsub} gives the probability for $m_{T}$ and $m_{T}$ events in  apertures $T$ and $S$, given the expectation values  $\bar{m}_{S}$ and $\bar{\mu}_{B}$~--- i.e., $P_{m_{T},m_{R}}(\bar{m}_{S},\bar{\mu}_{B}) = P(m_{T},m_{R}\, | \, \bar{m}_{S},\bar{\mu}_{B})$.
From this, one constructs a probability density function describing the (normalized) likelihood for the expectation values, given the observed distribution of events~--- i.e., $p(m_{S},\mu_{B} \, | \, m_{T},m_{R})$.
In order to facilitate this construction, we rewrite Equation~\ref{e:pmsub}, after slightly redefining variables and constant coefficients:
\begin{eqnarray}
P(m_{T},m_{R}\, | \, \nu_{S},\nu_{B}) & = & \frac{(\psi_{T}\nu_{S} + \omega_{T}\nu_{B})^{m_{T}}}{m_{T}!} \frac{(\psi_{R}\nu_{S} + \omega_{R}\nu_{B} )^{m_{R}}}{m_{R}!} \ e^{-(\nu_{S} + \nu_{B})} \ . \label{e:pm-nu}
\end{eqnarray}
The new variables are the expectation value for the number of source events in both apertures combined~--- $\nu_{S} \equiv \Psi_{U} m_{S} = (\Psi_{T}+\Psi_{R}) m_{S}$~--- and
the expectation value for the number of background events in both apertures combined~--- $\nu_{B} \equiv \Omega_{U} \mu_{B} = (\Omega_{T}+\Omega_{R}) \mu_{B}$.
The new (constant, predetermined) coefficients are the expected fractions of total source events in apertures $T$ and $R$~--- 
$\psi_{T} \equiv \Psi_{T}/\Psi_{U} = \Psi_{T}/(\Psi_{T}+\Psi_{R})$ and  $\psi_{R} \equiv \Psi_{R}/\Psi_{U} = \Psi_{R}/(\Psi_{T}+\Psi_{R})$, respectively~--- 
and the expected fractions of total background events in apertures $T$ and $R$~--- 
$\omega_{T} \equiv \Omega_{T}/\Omega_{U} = \Omega_{T}/(\Omega_{T}+\Omega_{R})$ and  $\omega_{R} \equiv \Omega_{R}/\Omega_{U} = \Omega_{R}/(\Omega_{T}+\Omega_{R})$, respectively.
Thus, $\psi_{T}+\psi_{R}=1$, so that $\psi_{T}$ or $\psi_{R}$ is the probability that a given source event occurs in aperture $T$ or $R$, respectively.
Analogously, $\omega_{T}+\omega_{R}=1$, so that $\omega_{T}$ or $\omega_{R}$ is the probability that a given background event occurs in aperture $T$ or $R$, respectively.
Expanding Equation~\ref{e:pm-nu} in terms of a double binomial series, we obtain
\begin{eqnarray}
P(m_{T},m_{R}\, | \, \nu_{S},\nu_{B}) & = & \sum_{i=0}^{m_{T}} \sum_{j=0}^{m_{R}} \frac{{\psi_{T}}^{i}\, {\psi_{R}}^{m_{R}-j}}{i! (m_{R}-j)!} \, \frac{{\omega_{R}}^{j}\, {\omega_{T}}^{m_{T}-j}}{j! (m_{T}-i)!} \nonumber\\
& & \times\ {\nu_{S}}^{i+m_{R}-j}\, {\nu_{B}}^{j+m_{T}-i}\ e^{-(\nu_{S} + \nu_{B})} \ . \label{e:pm-sum}
\end{eqnarray}

Dividing Equation~\ref{e:pm-sum} by the partition function $Z(m_{T}, m_{R})$~--- 
equivalent to normalizing the \cite[unweighted, cf.][]{kra91} integral of $P(m_{T},m_{R}\, | \, \nu_{S},\nu_{B})$ over all possible values $(0, \infty)$ of $\nu_{S}$ and $\nu_{B}$~--- 
we derive the desired probability density function:
\begin{eqnarray}
p(\nu_{S},\nu_{B} \, | \, m_{T},m_{R}) & = &
\frac{P(m_{T},m_{R}\, | \, \nu_{S},\nu_{B})}{Z(m_{T}, m_{R})} \nonumber \\
 & = & \frac{1}{Z(m_{T}, m_{R})}\ \sum_{i=0}^{m_{T}} \sum_{j=0}^{m_{R}} \frac{{\psi_{T}}^{i}\, {\psi_{R}}^{m_{R}-j}}{i! (m_{R}-j)!} \, \frac{{\omega_{R}}^{j}\, {\omega_{T}}^{m_{T}-j}}{j! (m_{T}-i)!} \nonumber\\  & & \times\ {\nu_{S}}^{i+m_{R}-j}\, {\nu_{B}}^{j+m_{T}-i}\ e^{-(\nu_{S} + \nu_{B})} \ , \label{e:pdf-sb}
\end{eqnarray}
where the partition function
\begin{eqnarray}
Z(m_{T}, m_{R}) & = & \sum_{i=0}^{m_{T}} \sum_{j=0}^{m_{R}} \frac{(i+m_{R}-j)!}{i! (m_{R}-j)!}\, {\psi_{T}}^{i}\, {\psi_{R}}^{m_{R}-j} \ \frac{(j+m_{T}-i)!}{j! (m_{T}-i)!}\, {\omega_{R}}^{j}\, {\omega_{T}}^{m_{T}-i} \ . \label{e:partition}
\end{eqnarray}
Note that the partition function (Equation~\ref{e:partition}) is the {\em a priori} probability that, given $m_{U} = m_{T}+m_{R}$ total (source and background) events, $m_{T}$ and $m_{R}$ events occur in apertures $T$ and $R$, respectively.

We may integrate the probability density $p(\nu_{S},\nu_{B} \, | \, m_{T},m_{R})$ from Equations~\ref{e:pdf-sb} and \ref{e:partition} to constrain jointly the values of $\nu_{S}$ and $\nu_{B}$ at a specified confidence level ${\cal C}$.
Alternatively, we may constrain either parameter individually, after integrating over the other's full range $(0, \infty)$.
Thus, the probability density for the expectation value $\nu_{S}$ of the number of source events in the combined aperture ($U \equiv T \cup R$), without regard to the value of $\nu_{B}$, is
\begin{eqnarray}
p(\nu_{S} \, | \, m_{T},m_{R}) & = & \frac{1}{Z(m_{T}, m_{R})}\ \sum_{i=0}^{m_{T}} \sum_{j=0}^{m_{R}} \frac{(j+m_{T}-i)!}{j! (m_{T}-i)!}\, {\omega_{R}}^{j}\, {\omega_{T}}^{m_{T}-i} \nonumber \\
& & \times\ \frac{{\nu_{S}}^{i+m_{R}-j}\ e^{-\nu_{S}}}{i! (m_{R}-j)!} \, {\psi_{T}}^{i}\, {\psi_{R}}^{m_{R}-j}\,  \ , \label{e:pdf-s}
\end{eqnarray}
For example, to establish an upper limit to the expectation value  $\bar{\nu}_{S}$ (without regard to $\nu_{B}$) at a confidence level ${\cal C}$, one solves
\begin{eqnarray}
1-{\cal C}(\bar{\nu}_{S} < \nu_{S} \, | \, m_{T},m_{R}) & = &  \int_{\nu_{S}}^{\infty} p(\nu_{S}' \, | \, m_{T},m_{R}) \ d\nu_{S}' \nonumber \\
& = & \frac{1}{Z(m_{T}, m_{R})}\ \sum_{i=0}^{m_{T}} \sum_{j=0}^{m_{R}} \frac{(j+m_{T}-i)!}{j! (m_{T}-i)!}\, {\omega_{R}}^{j}\, {\omega_{T}}^{m_{T}-i} \nonumber \\
& & \times\ \frac{\Gamma(i+m_{R}-j+1, \nu_{S})}{i! (m_{R}-j)!} \, {\psi_{T}}^{i}\, {\psi_{R}}^{m_{R}-j}\,  \ , \label{e:conf-s}
\end{eqnarray}
where $\Gamma(n+1, \nu)$ is the (upper) incomplete gamma function.

In the special case that $\psi_{T} \rightarrow 1$ and $\psi_{R} \rightarrow 0$~--- 
i.e., the expected fraction of source events in the reference aperture is negligible~--- 
the double sum reduces to the single sum
\begin{eqnarray}
1-{\cal C}(\bar{\nu}_{S} < \nu_{S} \, | \, m_{T},m_{R}) & \stackrel{\psi_{R} \rightarrow 0}{\longrightarrow} & 
\frac{1}{Z(m_{T}, m_{R})}\ \sum_{i=0}^{m_{T}} \frac{(m_{R}+m_{T}-i)!}{m_{R}! (m_{T}-i)!}\, {\omega_{R}}^{m_{R}}\, {\omega_{T}}^{m_{T}-i} \nonumber \\
& & \times\ \frac{\Gamma(i+1, \nu_{S})}{i!} \nonumber \\
& = & \frac{1}{Z(m_{T}, m_{R})}\ \sum_{k=0}^{m_{T}} \frac{(m_{R}+k)!}{m_{R}! k!}\, {\omega_{R}}^{m_{R}}\, {\omega_{T}}^{k} \nonumber \\
& & \times\ \frac{\Gamma(m_{T}-k+1, \nu_{S})}{(m_{T}-k)!} \ , \label{e:conf-s0}
\end{eqnarray}
The partition function also reduces to a single sum~--- namely,
\begin{eqnarray}
Z(m_{T}, m_{R}) & \stackrel{\psi_{R} \rightarrow 0}{\longrightarrow} & \sum_{i=0}^{m_{T}}  \ \frac{(m_{R}+m_{T}-i)!}{m_{R}!\, (m_{T}-i)!}\, {\omega_{R}}^{m_{R}}\, {\omega_{T}}^{m_{T}-i} \nonumber \\
& = & \sum_{k=0}^{m_{T}}  \ \frac{(m_{R}+k)!}{m_{R}!\, k!} \, {\omega_{R}}^{m_{R}}\, {\omega_{T}}^{k} \ . \label{e:partition0}
\end{eqnarray}

\clearpage

\begin{deluxetable}{lcccccccc}
\tabletypesize{\scriptsize}
\tablewidth{0pc}
\tablecaption{General properties of of GD~356 and other single white dwarfs. \label{t:prop}}
\tablehead{\multicolumn{1}{c}{(1)} & (2) & (3) & (4) & (5) & (6) & (7) & (8) & (9)}
\startdata
Name & RA(J2000)$^a$ & Dec(J2000)$^a$ & $\mu_{N}\, ^b$ & $\mu_{W}\, ^b$ & $D\, ^c$ & Spectral & ${T_{\rm eff}}^e$ & $B\, ^e$ \\
 & $\ ^{\rm h}\: \ ^{\rm m}\ \ \ ^{\rm s}\: $\ \ \ \  & \ $\ \arcdeg\: \ \ \arcmin\: \ \ \arcsec$\ \ \  & \arcsec/y & \arcsec/y & pc & type$^d$ & K & MG \\ \hline\\[-2ex]
LHS 1038 & 00 12 14.80 & +50 25 21.4 & $-$0.456 & $-$0.548 & 11.0 & DA8 & 6400 & 0.09\\
G~99-47  & 05 56 25.47 & +05 21 48.6 & $-$0.446 & $-$0.918 & 8.0  & DAP9 & 5600 & 27\\
GD~90    & 08 19 46.38 & +37 31 28.1 & $-$0.112 & $-$0.100 & 50   & DAH5 & 11000 & 10\\
G~195-19 & 09 15 56.23 & +53 25 24.9 & $-$1.080 & $-$1.116 & 10.3 & DCP7 & 8000 & 100\\
GD~356   & 16 40 57.16 & +53 41 09.6 & $-$0.118 & $-$0.186 & 21.1 & DAH & 7500 & 14\\
GD~358   & 16 47 18.39 & +32 28 32.9 & $-$0.166 & $+$0.026 & 36.6 & DBV2 & 24000 & 0.0013\\
2316+123 & 23 18 45.10 & +12 36 02.9 & $+$0.102 & $-$0.010 & 40 & DAp & 11800 & 56
\enddata \vspace{-0.125in}
\tablecomments{\\
$^a$ J2000 coordinates are from the 2MASS All-Sky Catalog of Point Sources \citep{skr06} for epoch 2000.00.\\
$^b$ J2000 proper-motion components $\mu_{N}$ and $\mu_{W}$ are from the Whole-Sky USNO-B1.0 Catalog \citep{mon03}.\\
$^c$ Distances are from the Yale Catalog of Trigonometric Parallaxes \citep{van95}, except (non-parallax) estimates for GD~90 and 2316+123 \citep{mus95}.\\
$^d$ Spectral types are from the Villanova Catalog of Spectroscopically Identified White Dwarfs \citep{mcc99}.\\
$^e$ Surface effective temperatures and magnetic fields are from \citet{jor01} and references therein, except for GD~358 \citep{win94}.}
\end{deluxetable}

\begin{deluxetable}{lccccccccccc}
\tabletypesize{\scriptsize}
\tablewidth{0pc}
\tablecaption{X-ray observations of GD~356 and other single white dwarfs. \label{t:x-data}}
\tablehead{\multicolumn{1}{c}{(1)} & (2) & (3) & (4) & (5) & (6) & (7) & (8) & (9) & (10) & (11) & (12)}
\startdata
Name & Obs.$^a$ & Epoch$^b$ & Time & ${m_{T}}^c$ & ${\Omega_{T}}^c$ & ${m_{R}}^c$ & ${\Omega_{R}}^c$ & ${{\cal C}_{\rm det}}^d$ & ${\bar{m}_{S}}\!^e$ & log[$L_{X}$]$^f$ & log[$n_{0}$]$^g$\\
 & & $t_{\rm obs}\!-\!2000$ & ks &  & $\arcsec^2$ &  & $\arcsec^2$ & \% &  & (cgs) & (cgs)\\ \hline\\[-2ex]
LHS 1038 & \Chandra & $+$0.97 & 5.88 & 0   & 3.14   & 2,029 & 234,000 &  0 & $<$5.0  & $<$25.95 & $<$11.72 \\
G~99-47  & \ROSAT   & $-$7.75 & 9.28 & 44  & 25,500 & 85    & 45,200  & 30 & $<$23.4 & $<$26.21 & $<$11.85 \\
GD~90    & \ROSAT   & $-$7.75 & 8.56 & 91  & 25,500 & 157   & 45,200  & 57 & $<$40.0 & $<$28.07 & $<$12.78 \\
G~195-19  & \ROSAT  & $-$7.69 & 6.99 & 90  & 25,500 & 195   & 45,200  &  5 & $<$23.9 & $<$26.57 & $<$12.03 \\
GD~356   & \Chandra & $+$5.39 & 31.8 & 0   & 3.14   & 5,437 & 216,000 &  0 & $<$5.0  & $<$25.78 & $<$11.64 \\
GD~356   & \ROSAT   & $-$7.99 & 28.2 & 234 & 25,500 & 413   & 45,200  & 52 & $<$60.5 & $<$26.99 & $<$12.24 \\
GD~358   & \Chandra & $+$1.79 & 4.88 & 0   & 3.14   & 2,402 & 233,000 &  0 & $<$5.0  & $<$27.06 & $<$12.28 \\
2316+123 & \ROSAT   & $-$8.08 & 9.12 & 44  & 25,500 & 61    & 45,200  & 88 & $<$37.7 & $<$27.83 & $<$12.66
\enddata \vspace{-0.125in}
\tablecomments{\\
$^a$ \Chandra\ observations used the ACIS-S instrument; \ROSAT\ observations used the PSPC.\\
$^b$ Proper-motion adjustment of coordinates requires epoch of X-ray observation relative to epoch of catalogued position.\\
$^c$ X-ray observation detected $m_{T}$ and $m_{R}$ events in Target $T$ and Reference $R$ apertures, of measure $\Omega_{T}$ and $\Omega_{R}$, respectively.\\
$^d$ Source-detection confidence is the probability that the background contributes fewer than the observed number of events in the Target aperture. (Equation~\ref{e:src-conf})\\
$^e$ Upper limit to the expectation value of 0.1--2.4-keV source events is at (3-$\sigma$) 99.7\%\ confidence. (Equation~\ref{e:conf-s})\\
$^f$ Calculated X-ray luminosity assumes thermal bremsstrahlung at $kT = 1$~keV.\\
$^g$ Calculated electron density at coronal base assumes thermal bremsstrahlung at $kT = 1$~keV, from a fully ionized plasma with $n_{\rm He}/n_{\rm H}=0.1$, above an opaque white dwarf of radius $0.0105\, R_{\odot} = 7.34\!\times\!10^{8}$~cm and surface gravity log[$g$(cgs)] = 8.}
\end{deluxetable}

\clearpage

\begin{deluxetable}{crrrrr}
\tabletypesize{\scriptsize}
\tablewidth{0pc}
\tablecaption{\Chandra-detected sources in the GD-356 field. \label{t:field_gd356}}
\tablehead{(1) & \multicolumn{1}{c}{(2)} & \multicolumn{1}{c}{(3)} & \multicolumn{1}{c}{(4)} & \multicolumn{1}{c}{(5)} & \multicolumn{1}{c}{(6)}} 
\startdata
 Source & \multicolumn{1}{c}{RA(J2000)} & \multicolumn{1}{c}{Dec(J2000)} & \multicolumn{1}{c}{${\theta_{\rm ext}}^a$} & \multicolumn{1}{c}{${m_{D}}^b$} &  \multicolumn{1}{c}{${\theta_{99}}^c$} \\
 & \multicolumn{1}{c}{$^{\rm h}\: \ ^{\rm m}\ \ ^{\rm s}\: $\ \ \ \ } & \multicolumn{1}{c}{\ $\ \arcdeg\: \ \ \arcmin\: \ \ \arcsec$\ \ \ }
 & \multicolumn{1}{c}{$\arcsec$} & & \multicolumn{1}{c}{$\arcsec$}  \\ \hline\\[-2ex]
      1 &   16 40 45.029 &   +53 44 47.38 &    2.7 &      8.9 &  1.25 \\
      2 &   16 40 45.299 &   +53 45 13.15 &    3.1 &     13.6 &  1.18 \\
      3 &   16 40 53.706 &   +53 44 42.50 &    2.4 &      9.2 &  1.15 \\
      4 &   16 40 55.545 &   +53 40 25.24 &    1.3 &     22.0 &  0.69 \\
      5 &   16 40 56.104 &   +53 39 18.33 &    1.7 &     25.9 &  0.73 \\
      6 &   16 40 59.258 &   +53 41 59.41 &    1.2 &     36.4 &  0.66 \\
      7 &   16 41 00.438 &   +53 42 03.31 &    1.3 &     10.2 &  0.78 \\
      8 &   16 41 04.242 &   +53 40 20.98 &    1.6 &      7.3 &  0.94 \\
      9 &   16 41 05.997 &   +53 43 18.82 &    2.0 &     21.0 &  0.80 \\
     10 &   16 41 06.757 &   +53 37 52.85 &    3.3 &     30.6 &  0.94 \\
     11 &   16 41 07.650 &   +53 45 27.75 &    3.7 &     17.8 &  1.23 \\
     12 &   16 41 10.745 &   +53 44 36.43 &    3.2 &     16.5 &  1.13 \\
     13 &   16 41 13.050 &   +53 41 57.24 &    2.2 &     12.9 &  0.97 \\
     14 &   16 41 14.806 &   +53 41 41.18 &    2.4 &     12.6 &  1.02 \\
     15 &   16 41 15.481 &   +53 44 10.91 &    3.4 &   4559.0 &  0.61 \\
     16 &   16 41 16.876 &   +53 42 56.29 &    2.9 &    118.6 &  0.69 \\
     17 &   16 41 19.134 &   +53 44 11.29 &    3.9 &     32.1 &  1.03 \\
     18 &   16 41 21.033 &   +53 40 54.25 &    3.3 &     17.6 &  1.13 \\
     19 &   16 41 24.066 &   +53 41 48.86 &    3.8 &      8.2 &  1.71 \\
     20 &   16 41 30.445 &   +53 41 18.79 &    5.0 &     27.8 &  1.30 \\
     21 &   16 41 32.519 &   +53 36 40.09 &    8.4 &     17.8 &  2.49 \\
     22 &   16 41 33.184 &   +53 43 05.51 &    5.9 &     12.7 &  2.10 \\
     23 &   16 41 37.639 &   +53 39 57.65 &    6.9 &      8.5 &  2.93 \\
\enddata \vspace{-0.125in}
\tablecomments{\\
$^a$ The extraction radius demarks the detect cell for collecting X-ray counts.\\
$^b$ An  ACIS-S3 observation (ObsID~4484) acquired these detect-cell counts in 31.8~ks.\\
$^c$ This radius encloses the true position of the detected source at 99\% confidence.}
\end{deluxetable}

\clearpage

\begin{deluxetable}{crrrrr}
\tabletypesize{\scriptsize}
\tablewidth{0pc}
\tablecaption{\Chandra-detected sources in the LHS-1038 field. \label{t:field_lhs1038}}
\tablehead{(1) & \multicolumn{1}{c}{(2)} & \multicolumn{1}{c}{(3)} & \multicolumn{1}{c}{(4)} & \multicolumn{1}{c}{(5)} & \multicolumn{1}{c}{(6)} }
\startdata
 Source & \multicolumn{1}{c}{RA(J2000)} & \multicolumn{1}{c}{Dec(J2000)} & \multicolumn{1}{c}{${\theta_{\rm ext}}^a$} & \multicolumn{1}{c}{${m_{D}}^b$} &  \multicolumn{1}{c}{${\theta_{99}}^c$} \\
 & \multicolumn{1}{c}{$^{\rm h}\: \ ^{\rm m}\ \ ^{\rm s}\: $\ \ \ \ } & \multicolumn{1}{c}{\ $\ \arcdeg\: \ \ \arcmin\: \ \ \arcsec$\ \ \ }
 & \multicolumn{1}{c}{$\arcsec$} & & \multicolumn{1}{c}{$\arcsec$}  \\ \hline\\[-2ex]
      1 &    0 12 05.742  &  +50  28 17.79 &   2.8 &    11.3 &   1.16 \\
      2 &    0 12 08.128  &  +50  30 16.70 &   4.6 &    10.1 &   1.86 \\
      3 &    0 12 10.118  &  +50  28 30.63 &   2.7 &    60.9 &   0.73 \\
      4 &    0 12 31.360  &  +50  32 16.93 &   8.1 &    35.5 &   1.75 \\
\enddata \vspace{-0.125in}
\tablecomments{\\
$^a$ The extraction radius demarks the detect cell for collecting X-ray counts.\\
$^b$ An  ACIS-S3 observation (ObsID~1864) acquired these detect-cell counts in 5.88~ks.\\
$^c$ This radius encloses the true position of the detected source at 99\% confidence.}
\end{deluxetable}

\begin{deluxetable}{crrrrr}
\tabletypesize{\scriptsize}
\tablewidth{0pc}
\tablecaption{\Chandra-detected sources in the GD-358 field. \label{t:field_gd358}}
\tablehead{(1) & \multicolumn{1}{c}{(2)} & \multicolumn{1}{c}{(3)} & \multicolumn{1}{c}{(4)} & \multicolumn{1}{c}{(5)} & \multicolumn{1}{c}{(6)} }
\startdata
 Source & \multicolumn{1}{c}{RA(J2000)} & \multicolumn{1}{c}{Dec(J2000)} & \multicolumn{1}{c}{${\theta_{\rm ext}}^a$} & \multicolumn{1}{c}{${m_{D}}^b$} &  \multicolumn{1}{c}{${\theta_{99}}^c$} \\
 & \multicolumn{1}{c}{$^{\rm h}\: \ ^{\rm m}\ \ ^{\rm s}\: $\ \ \ \ } & \multicolumn{1}{c}{\ $\ \arcdeg\: \ \ \arcmin\: \ \ \arcsec$\ \ \ }
 & \multicolumn{1}{c}{$\arcsec$} & & \multicolumn{1}{c}{$\arcsec$}  \\ \hline\\[-2ex]
      1 &   16 46 55.620 & +32 29 49.17 &   4.8 &     6.9 &  2.29 \\
      2 &   16 46 58.650 & +32 29 30.20 &   3.9 &     9.6 &  1.64 \\
      3 &   16 47 07.418 & +32 30 50.62 &   3.0 &    25.1 &  0.95 \\
      4 &   16 47 08.044 & +32 25 17.23 &   2.7 &    17.2 &  0.99 \\
      5 &   16 47 13.515 & +32 32 03.03 &   3.5 &    20.6 &  1.12 \\
      6 &   16 47 17.610 & +32 33 15.56 &   4.8 &    25.3 &  1.31 \\
      7 &   16 47 18.628 & +32 27 51.63 &   1.1 &    51.1 &  0.63 \\
      8 &   16 47 18.645 & +32 29 16.22 &   1.4 &    13.3 &  0.76 \\
      9 &   16 47 26.652 & +32 27 30.17 &   1.4 &     8.4 &  0.85 \\
     10 &   16 47 27.949 & +32 27 37.51 &   1.5 &    13.0 &  0.80 \\
\enddata \vspace{-0.125in}
\tablecomments{\\
$^a$ The extraction radius demarks the detect cell for collecting X-ray counts.\\
$^b$ An  ACIS-S3 observation (ObsID~1865) acquired these detect-cell counts in 4.88~ks.\\
$^c$ This radius encloses the true position of the detected source at 99\% confidence.}
\end{deluxetable}

\clearpage

\begin{figure}[htbp]
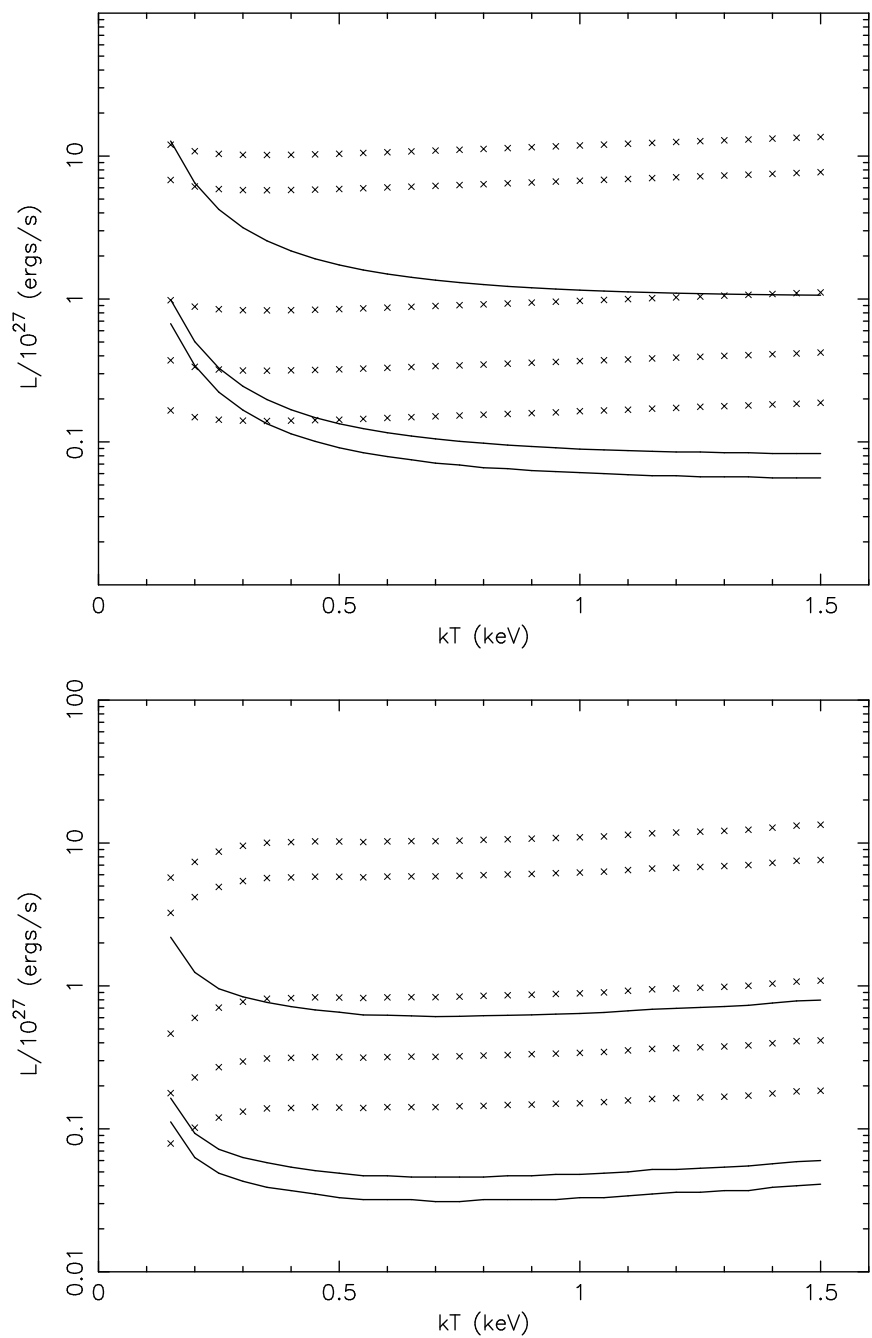

\center{\epsfig{file=fig1a.ps, angle=-90, width=4.5in}}
\center{\epsfig{file=fig1b.ps, angle=-90, width=4.5in}}
\caption{
The $99.7$\%-confidence upper limits to total X-ray luminosity versus temperature (upper panel) for bremsstrahlung and (lower panel) for MEKAL models.
From lowest to highest, limits based upon \Chandra\ data (solid lines) are for GD~356, LHS~1038, and GD~358; those based upon \ROSAT\ data ($\times$ markers) are for G~99-47, G~195-19, GD~356, 2316+123, and GD~90.
\label{f:luminosity}}
\end{figure}

\clearpage

\begin{figure}[htbp]
\center{\epsfig{file=fig2a.ps, angle=-90, width=4.5in}}
\center{\epsfig{file=fig2b.ps, angle=-90, width=4.5in}}
\caption{
The $99.7$\%-confidence upper limits to coronal electron density versus temperature (upper panel) for bremsstrahlung and (lower panel) for MEKAL models.
From lowest to highest, limits based upon \Chandra\ data (solid lines) are for GD~356, LHS~1038, and GD~358; those based upon \ROSAT\ data ($\times$ markers) are for G~99-47, G~195-19, GD~356, 2316+123, and GD~90.
\label{f:density}}
\end{figure}

\begin{figure}[htbp]
\center{\epsfig{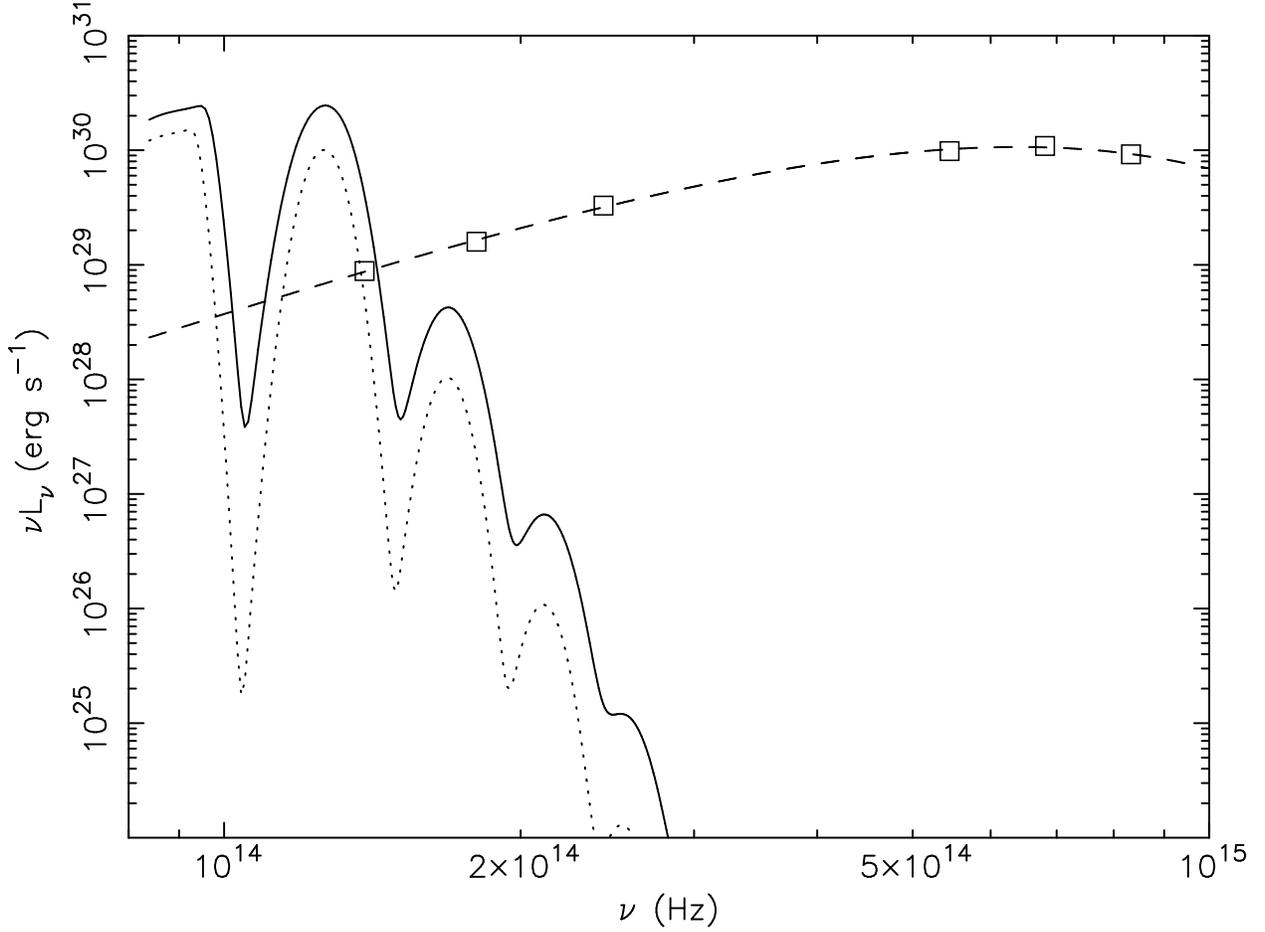}}
\caption{Emergent cyclotron spectrum from a hypothesized hot corona around GD~356, in comparison to a 7840-K photospheric spectrum (dashed line) and photometric data (square markers).
Calculations assume a 15-MG field at a mean viewing angle $\theta = \arctan (0.5) = 26.6\degr$ and coronal parameters discussed in the text (\S \ref{s:chandra_obs}), for a (coronal-base) electron density $n_{0} = 1\!\times\!10^{11}$~cm$^{-3}$ and temperature $T = 1.5$~keV (solid line) or 1.0~keV (dotted line).
\label{f:cyclotron}}
\end{figure}

\begin{figure}[htbp]
\center{\epsfig{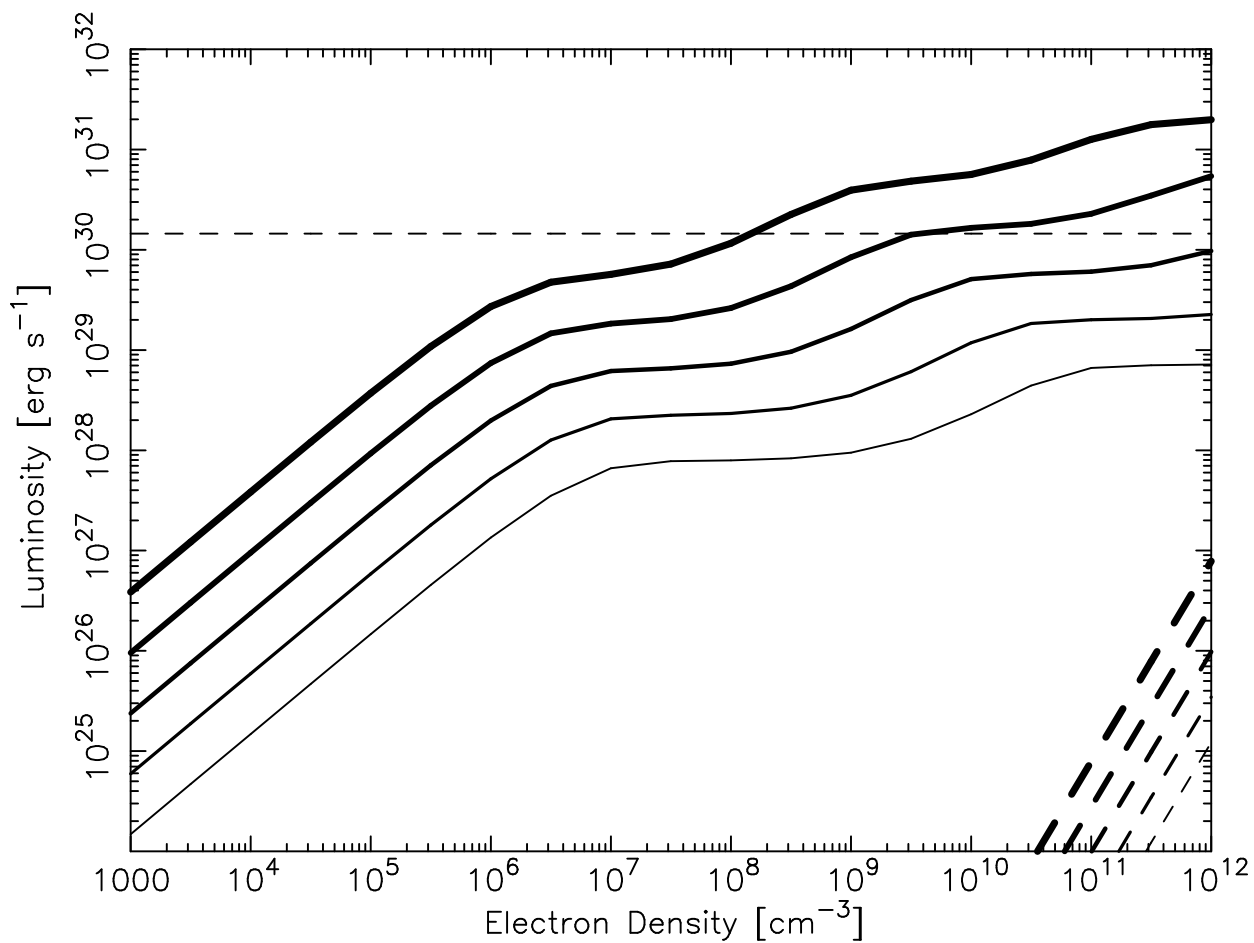}}
\caption{
Contributors to the luminosity of a hypothetical corona around the magnetic white dwarf GD 356, in comparison to its photospheric luminosity (horizontal, thin dashed line).
Solid lines denote the thermal cyclotron luminosity for $B =$~15~MG and $kT =$~0.125, 0.25, 0.5, 1.0, and 2.0~keV, in increasing thickness.
Dotted lines denote the thermal-bremsstrahlung luminosity for the same set of coronal temperatures.}
\label{f:power}
\end{figure}

\end{document}